# A Light-Driven Microgel Rotor

*Hang Zhang†, Lyndon Koens, Eric Lauga, Ahmed Mourran\*, Martin Möller*

Dr. Hang Zhang, Dr. Ahmed Mourran, Prof. Matin Möller
DWI Leibniz-Institute for Interactive Materials, RWTH Aachen University, Forckenbeckstr. 50, D-52056 Aachen, Germany
E-mail: mourran@dwi.rwth-aachen.de

Dr. Lyndon Koens
Department of Mathematics and Statistics, Macquarie University, 192 Balaclava Rd, Macquarie Park NSW 2113, Australia

Prof. Eric Lauga
Department of Applied Mathematics and Theoretical Physics, University of Cambridge, Wilberforce Road, Cambridge CB3 0WA, United Kingdom

† Current address: Department of Applied Physics, Aalto University, Puumiehenkuja 2, 02150 Espoo, Finland

**Abstract:** The current understanding of motility through body shape deformation of microorganisms and the knowledge of fluid flows at the microscale provides ample examples for mimicry and design of soft microrobots. In this work, a two-dimensional spiral is presented that is capable of rotating by non-reciprocal curling deformations. The body of the microswimmer is a ribbon consisting of a thermo-responsive hydrogel bilayer with embedded plasmonic gold nanorods. Such a system allows fast local photothermal heating and non-reciprocal bending deformation of the hydrogel bilayer under non-equilibrium conditions. We show that the spiral acts as a spring capable of large deformations thanks to its low stiffness, which is tunable by the swelling degree of the hydrogel and the temperature. Tethering the ribbon to a freely rotating microsphere enables rotational motion of the spiral by stroboscopic irradiation. The efficiency of the rotor is estimated using resistive force theory for Stokes flow. The present research demonstrates microscopic locomotion by the shape change of a spiral and may find applications in the field of microfluidics, or soft micro-robotics.

# 1. Introduction

Microorganisms are ubiquitous in our world and their locomotion in the form of swimming is an essential part of their lives. Prokaryotes (for example the model bacterium *Escherichia coli*) and eukaryotes (e.g., *Paramecium* and alga *Chlamydomonas*) swim to explore their chemical environment,[1,2] pursue food,[3] or escape predation.[4] Similarly, spermatozoa swim using flagella to spread progeny. Being small, microorganisms have developed swimming strategies that do not apply to the macroscopic world due to the high viscous drag experienced by the body at a length scale where the inertial forces are negligible.[5,6] The ratio between these forces defines the Reynolds number *Re*, which is of the order of $10^{-4}$ for microorganisms.[5] Two major mechanisms used by microorganism to swim and stir surrounding fluids are the beating of the cilia and rotating of the helical flagella, which are filaments capable of active bending deformation or rotation.[5–10] These filaments driven by internally generated forces work with viscous and elastic forces to generate propulsive thrusts.[11–13] The physical balance between elastic forces and viscous drag determines the swimming speed, or pumping frequency and performance.[14]

To our best knowledge, a microswimmer based on curling of a two-dimensional spiral is yet to be found in nature, though helical-shaped microorganisms are abundant, such as *spirilla* and spirochetes. Curling is a way to store elastic energy and is widely used in engineering, such as piezoelectric devices,[15] energy storage in clock springs, or 3D displacement in artificial muscles.[16] In this report, we present a novel type of microswimmer, i.e. a spiral-shaped microgel that is capable of rotating by non-reciprocal deformations. The body of the microswimmer is ribbon-shaped and consists of a crosslinked poly(*N*-isopropylacrylamide) (PNIPAM) hydrogel laden with gold nanorods (AuNRs). AuNRs are engineered to absorb photon energy in the near infrared and generate localized heat that triggers volume changes of the surrounding hydrogel.[17–

[19] Coated on one surface with a thin metal layer, the bilayer hydrogel ribbon is able to swell and curl in water along its length to form a two-dimensional spiral.[20-22] Upon heating, the solubility decrease of the polymer network causes the hydrogel layer to shrink, so that the spiral unwinds. We note here that the bending rigidity of the ribbon and its stiffness are greatly affected by the water content in it. The light-induced temperature-jumps give rise to mechanical response over a few hundred milliseconds, where the rate-limiting factor is the mass transport within the gel.[22-28] We hypothesize that photothermal heating creates a transient state, where an imbalance exists between the stresses defining the local and the mean curvature of the ribbon.[21] These elastic restoring forces of bending and stretching are counterbalanced by the viscous drag of the surrounding fluid, which sets the rotor in motion. Since the material is relatively soft, the curling dynamics depends strongly on the dissipation mechanisms.[21,29,30]

Unlike previous studies that focused on the end equilibrium states of the hydrogel volume-phase transition, we study the response dynamics upon short-lived stimuli here, an aspect that is less well-understood.[27,28,31]. We show that the light-driven microgel undergoes time-dependent deformations and provides mechanical work. The light energy in the near-infrared causes the gel's shape to change outside the thermodynamic equilibrium, and it is possible to probe the kinematics of the deformation by the frequency of the light's strobe.[17] The material is shaped by the swelling of the gel, reminiscent to the hydraulic force that erects the butterfly proboscis or that coils tendrils plants,[32,33] raising questions about the differences and similarity of the synthetic ribbon to those biological counterparts. In the following, the fabrication and light-driven motion of the spiral microrotor is demonstrated. Through quasi-static heating, we assessed the changes in curvature as a function of the misfit strain. A measure of the maximum speed would require linking the bending wave and stiffness of the microgel ribbon with respect to the frequency of cyclic irradiation,[7] an issue not addressed here but will be pursued in the near future. Finally, the efficiency of the system is estimated based on the resistive force theory.

## 2. Results and discussion

### 2.1. Thermoresponsive property of a two-dimensional spiral

The fabrication of the two-dimensional spiral microgel is illustrated schematically in **Figure 1** (a-e) along with an optical micrograph showing the resultant spiral microgel in water at 20 °C. Details of the process can be found in the experimental section. Briefly, a non-wetting soft mold is replicated from a silicon master (**Figure S1**) made by photolithography with desired dimensions. Isolated microgel particles containing AuNRs can be then molded inside the soft template by photo-polymerization with highly defined shape and composition.[20,34] After polymerization, the top surface of the microgels is sputter-coated by a gold skin layer, after which the microgels are released into water by a transfer process. Upon swelling in water, the expansion of the gel layer is restricted by the gold skin, which resists stretching. It is therefore a bilayer ribbon in which one layer expands more relative to the other upon actuation.[35] Consequently, swelling of the hydrogel causes differential strain that leads to bending of the bilayer. Depending on the initial geometry of the ribbon, different structures can be formed such as a helix or two-dimensional spiral.[20] In this report, we focus on the ribbon with an as-prepared size of $(5 \times 10 \times 200)$ µm$^3$, which bends and forms a two-dimensional spiral upon swelling in water.

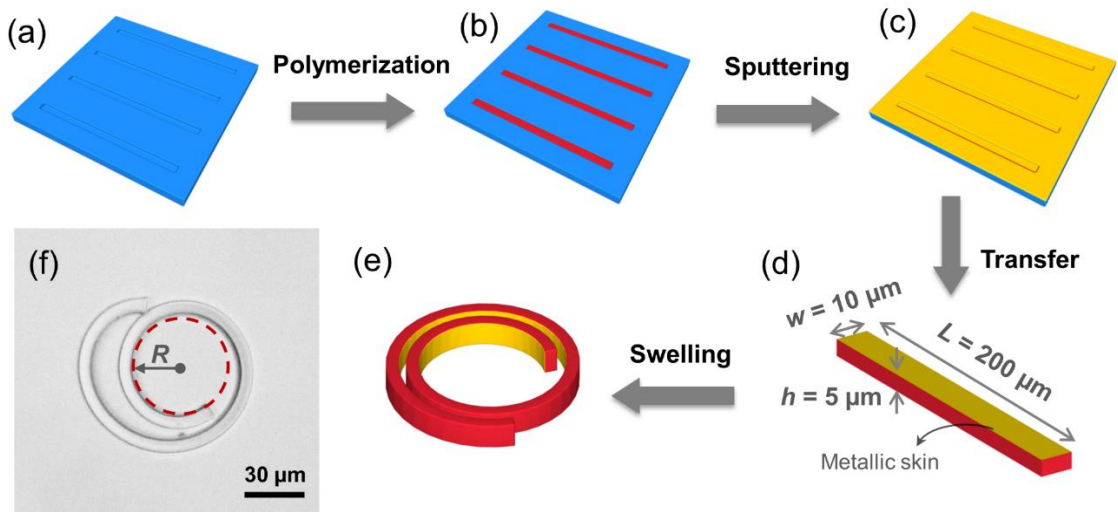

**Figure 1. Fabrication of bilayer spiral microgels.** (a) – (e) Schematic illustration of the fabrication of the hydrogel bilayer. (a) Perfluoropolyether (PFPE) mold replicated from a silicon master. (b) Microgels polymerized inside the mold. (c) Sputter-coating of a thin metallic skin on the microgel. (d) Dimension of the as-prepared microgel particle. (e) Microgel bilayer after swelling. (f) Optical micrograph of a spiral microgel at 20 °C. Inner radius $R$ is indicated by the dashed circle.

The thermo-response of the microgel bilayer was investigated by measuring the change of the curvature at different temperatures. **Figure 2** shows that as the temperature approaches the volume phase transition temperature of the PNIPAM network, the spiral unbends and nearly recovers its original straight shape close to. The diagram in Fig. 2b shows the straightening of the spiral with temperature, which increases the radius and decreases the length of the ribbon due to the de-swelling of the gel layer. This property can be understood by examining the dependence of the curvature on the relative strain normalized to the thickness, i.e., $\varepsilon/h$, where $\varepsilon = L - L_0$ ($L_0 = 200$ µm) represents the relative variation in the ribbon length and $h$ is the thickness (see Figure 2c). The dependence is rather linear, which is reminiscent of Timoshenko's bending bimetallic strip.[36] In that case the curvature is proportional to the misfit strain between the two layers and inversely proportional to the thickness of the strip. It is also worth noting that the slope is close to that of a helical microgel with a similar aspect ratio,[20] which emphasizes the role of geometry and hydrogel composition in defining the mechanical response of the hydrogel bilayer. The dramatic change of the curvature, especially near the volume-phase transition temperature, is an important feature of the bilayer system that can be utilized to perform mechanical work at the expense of a small change in the volume.

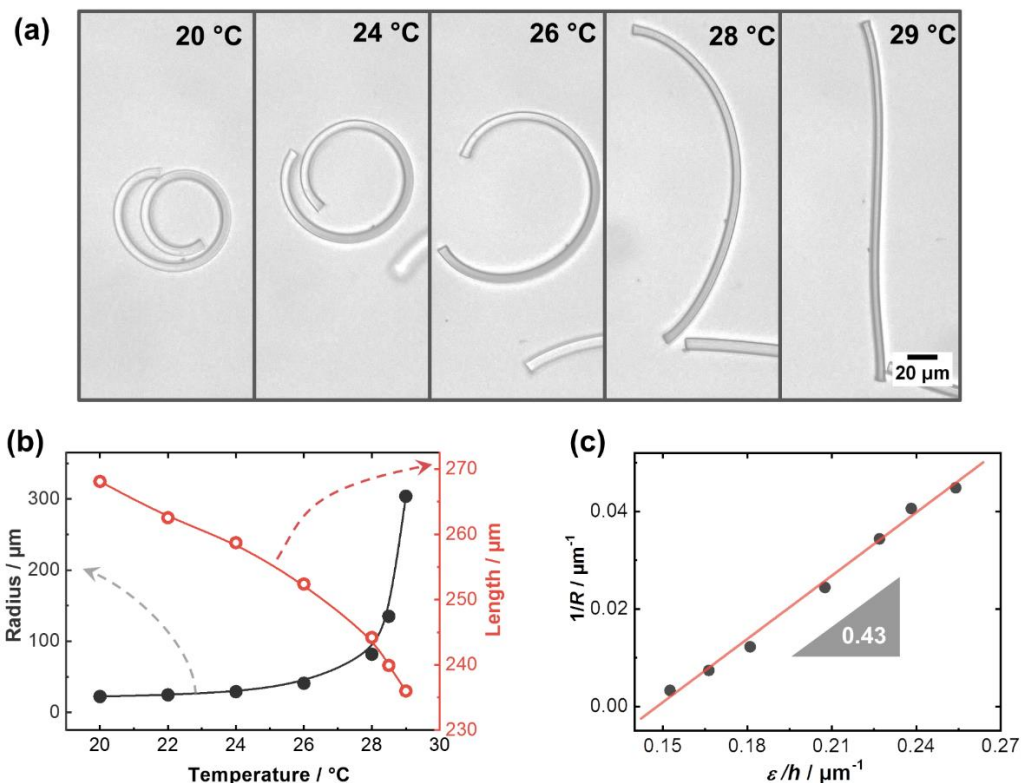

**Figure 2. Characterization of the spiral microgel at different temperatures.** (a) Optical images of the spiral at different temperatures. (b) The change of the spiral length and the radius with temperature. (c) Curvature (1/$R$) of the spiral versus strain $\varepsilon$ relative to the ribbon thickness $h$.

### 2.2. Non-reciprocal actuation via photothermal heating

To fabricate the microgel rotor, a silica microsphere (diameter ~ 10 µm) was attached to one end of the microgel as shown in the optical micrograph of **Figure 3c**, which was done by mixing the spiral microgels with silica particles and sorting out the microgel which was attached by the microsphere at one end. The spiral was then confined edge-on between two glass surfaces such that the microsphere can freely rotate without significant lateral displacement (**Figure S2, S3**).

Upon stroboscopic irradiation at 808 nm, the spiral underwent bending/unbending deformations as demonstrated in Figure 3a and Supporting Videos 1 - 4.

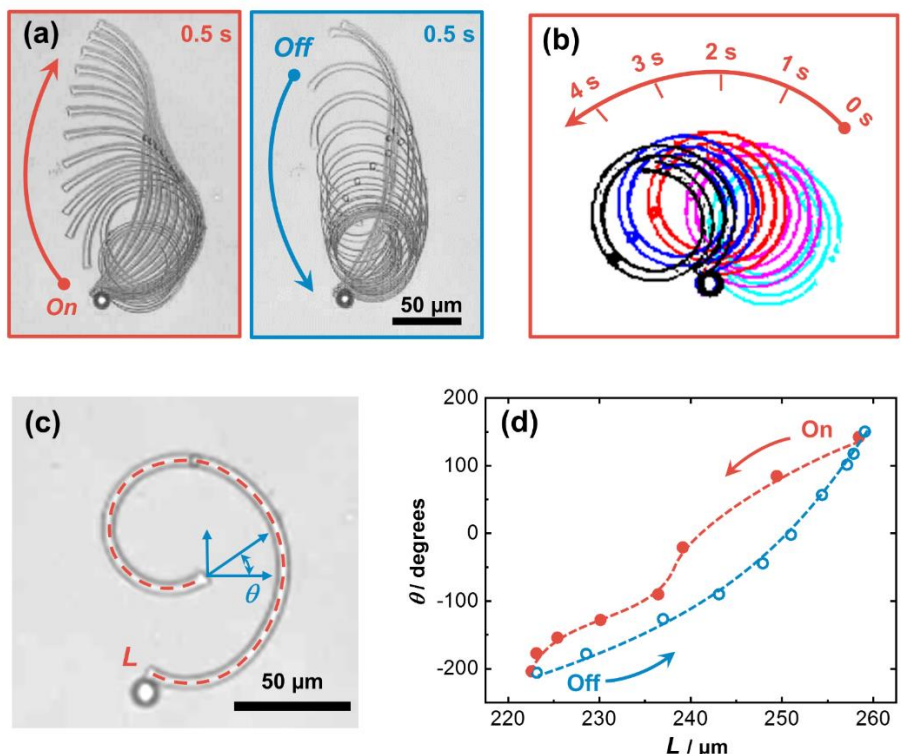

**Figure 3. Photothermal actuation of a spiral rotor tethered to a freely rotating sphere.** (a) Superimposed optical micrographs of the spiral rotor under stroboscopic irradiation. Left: on-time (0.5 s); right: off-time (0.5 s). Laser intensity: 1.7 W mm$^{-2}$ at 808 nm. (b) Superimposition of the thresholded outline of the spiral rotor under stroboscopic irradiation (0.5 s – 0.5 s). From right to left: at 0 s, 1 s, 2 s, 3 s, and 4 s. (c) Optical micrograph of the spiral rotor during recovery. The length of the center line ($L$, red-dashed line) and the tangential angle ($\theta$) are marked correspondingly. (d) Change of the $L$ vs. $\theta$ during one cycle of irradiation (0.5 s – 0.5 s). Arrows mark the direction of change, and dashed lines are drawn to guide the eye.

Figure 3a shows the shape-superimposition during an actuation cycle, i.e. irradiation for 500 ms and recovery (light off) for 500 ms. The period was chosen based on the characteristic time for the water to diffuse in and out of the microgel with a thickness of 5 μm, which is in the range from sub-second to a few seconds.[17,37,38] The spiral continuously unfurls to an almost straight shape within 0.5 s of irradiation as seen in the optical micrographs, and re-winds during the recovery stroke (0.5 s off). At the end of each irradiation cycle, the microgel recovers its spiral shape with an anti-clockwise rotation with respect to the initial position. The Reynolds number of the spiral microgel can be estimated to be around $10^{-2}$ using the width of the microgel as the characteristic length (10 µm) and a typical velocity of 1000 µm s$^{-1}$. The rotation of the spiral can be clearly observed after a few cycles as shown by the superimposed images taken at the beginning of each cycle in Figure 3b. The spiral exhibited a net anticlockwise rotation around the microsphere with an average speed of 22 degrees per cycle.

As shown in previous work, one necessary condition to obtain swimming at low Reynolds numbers is that the deformation path during one cycle should follow different paths in the shape space.[5–7,39] Indeed, it can be observed from the superimposition in Figure 3a that the deformation sequence during the recovery stroke is different from the power stroke, and the evolution of curvature differs as well. The kinematics of the spiral shows a close resemblance to the cilia in eukaryotic cells as demonstrated in literature.[40] The power stroke of a real cilium is characterized by the almost rigid rotation around its base, while the recovery stroke shows a large curvature to produce the non-reciprocal deformations, i.e. the non-identical sequence of shapes during one cycle under a time-reversal symmetry. Two geometric parameters were selected to characterize the deformation: the length of the center line ($L$) and the tangential angle ($\theta$) of the free end as illustrated in Figure 3c, see **Figure S4** for the image analysis.[41] The evolution of these two parameters is plotted in Figure 3d with the $L$ as X-axis and $\theta$ as Y-axis, demonstrating the time-independent deformation during one cycle. Unbending of the spiral structure under laser irradiation resulted in a decrease of the length from 258 µm to 222 µm and a decrease of $\theta$ from 142 to -203 degrees. The continuous decrease in $L$ is due to the shrinking of the gel, while the change in the $\theta$ indicates the local deformation of the spiral. The microgel contour length at the end of the irradiation is close to the length at 29 °C (as indicated in Figure 2b). In contrast, the recovery stroke (off-time) follows a distinct path different to that of the power-stroke. The steady increase in the length during the off-time is due to the swelling of the hydrogel layer upon cooling. Meanwhile, the winding of the spiral results in the continuous increase of $\theta$ to more than 150 degrees. This value is slightly larger than the starting point in the cycle due to the net rotation of the spiral after one cycle of actuation. The non-reciprocal deformation can be attributed to the combination of non-equilibrium actuation and bending of the bilayer.[20]

**Figure 4** shows the influence of irradiation period on the spiral net rotation. For a short irradiation period of 0.1 s – 0.1 s, the spiral microgel is not able to wind back due to the short recovery time. The actuation amplitude defined as the displacement of the free end before and after irradiation, is rather small in this case. Nevertheless, the asymmetry of the motion kinematics during an irradiation cycle is still distinguishable in Figure 4a with a net rotational motion of 1.6 degree s$^{-1}$ (see Figure 4d). Longer irradiation periods increase the motion amplitude and enhance the asymmetry in the motion kinematics as shown in Figure 4 b and c. Figure 4b shows a significantly larger amplitude as well as asymmetric kinematics compared to Figure 4a with an increased rotational speed, i.e. 16.7 degree s$^{-1}$. The longest irradiation period is 2 s in Figure 4c. Note that at the end of the irradiation, the ribbon is slightly bent in the opposite direction. This can be attributed to the shrinkage of the hydrogel causing the bilayer to bend in the opposite direction.[20] The angular velocity in this case is 8.8 degree s$^{-1}$. In Figure 4d, the angular velocity of rotation shows a maximum of 22 degree s$^{-1}$ at the $t_{on}$ of 0.5 s. This is in an interesting correlation with the amplitude of actuation, where the largest amplitude appears also at the $t_{on}$ of 0.5 s. At this irradiation period, the spiral microgel already reached a fully stretched state, so that the amplitude did not further increase upon longer irradiation. The angular velocity decreased at the $t_{on}$ of 1.0 s, even though the rotation per cycle is close to the 0.5 s.

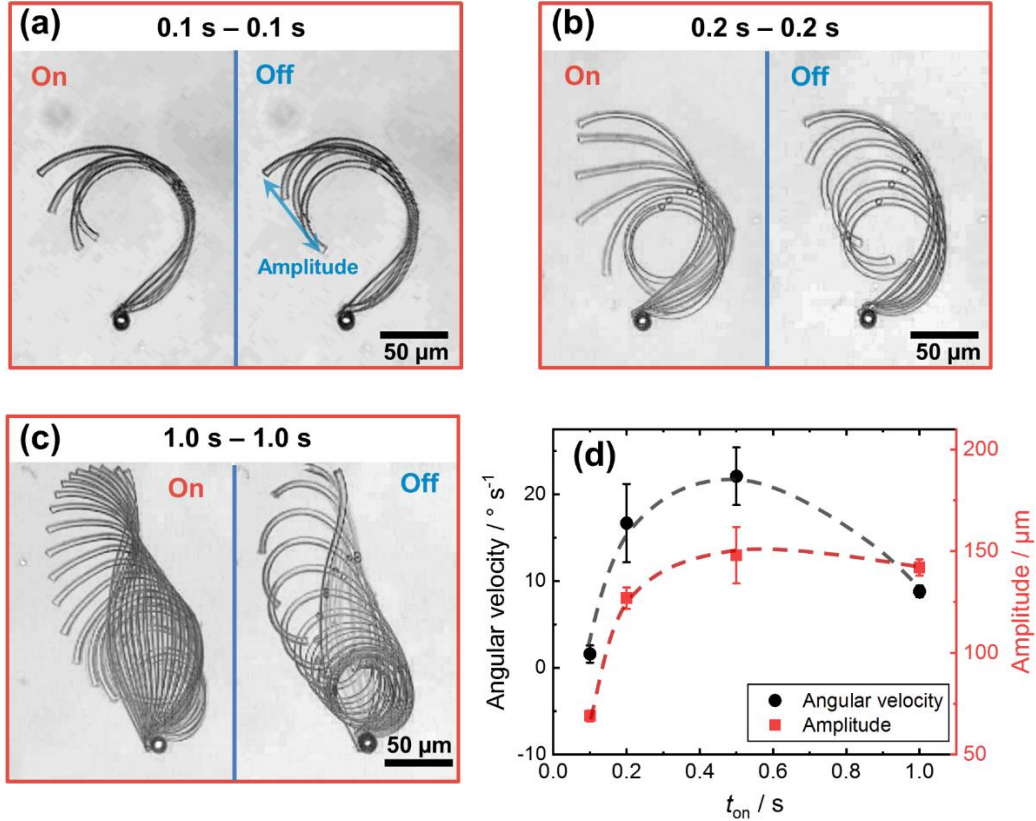

**Figure 4. Spiral rotor at different modulation frequency.** (a), (b), and (c) Superimposed optical micrographs of the spiral under different stroboscopic irradiation. On – off irradiations time from (a) to (c): 0.1 s – 0.1s, 0.2 s – 0.2 s, and 1.0 s - 1.0 s. Scale bar: 50 µm. Laser intensity: 1.7 W mm$^{-2}$ at 808 nm. The amplitude of actuation is marked in (a). (d) Dependency of angular velocity on the duration of on-time $t_{on}$. Each data point is averaged from five cycles of irradiation, and the error bar shows the standard deviation. Dashes lines are drawn to guide the eyes.

## 2.3. Theoretical estimation of power and efficiency

The efficiency of the deforming spiral rotor is determined by the ratio of the power required to rotate the body compared to the input power, therefore an efficient rotor would have a scalar value of 1. At microscopic scales, the drag from the fluid on the body is linearly related to the object's linear and angular velocity. As such the torque on the fluid and the power dissipation from the rotation can be written as

$$\tau = R_\tau \Omega, \tag{1}$$

$$P = \tau\Omega = R_\tau\Omega^2, \tag{2}$$

where $\tau$ is the torque on the fluid, $\Omega$ is the angular velocity, $P$ is the power, and $R_\tau$ is a linearity coefficient that depends on the geometry of the shape. For long thin objects in Stokes flow, $R_\tau$ can be estimated with a technique called resistive force theory.[42] This theory relates the force per unit length along the object, $f$, to the velocity of the filament at that location, $U$, through the equation

$$\mathbf{f} = [\zeta_t \hat{\mathbf{t}}\hat{\mathbf{t}} + \zeta_n(\mathbf{I} - \hat{\mathbf{t}}\hat{\mathbf{t}})] \cdot \mathbf{U}, \tag{3}$$

where $\hat{t}$ is the tangent vector to the centreline of the filament and $\zeta_t$ and $\zeta_n$ are drag coefficients for motion tangent and perpendicular to the centreline, respectively. This relationship is shown diagrammatically in **Figure 5**.

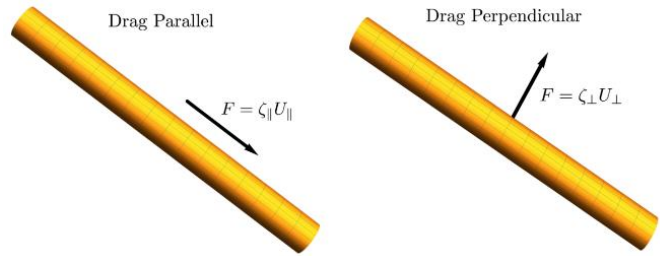

**Figure 5. Diagrammatic illustration of the drag parallel and perpendicular to the filament.**

These drag coefficients increase significantly when very close to walls. In this limit the flow is known as the lubrication flow and is governed by steep gradients in the velocity. In this lubrication limit the drag per unit length on a cylinder, from a single wall, moving normal to the axis is given by [43]

$$\zeta_{n,1\text{wall}} = \frac{2\pi\mu\sqrt{2r_b}}{\sqrt{\epsilon}}, \tag{4}$$

where µ is the viscosity of the fluid, $r_b$ is the radius of the filament and $\epsilon$ is the separation between the particle and the wall. Since the spiral rotor is closely confined between two walls, and lubrication forces are additive, the normal drag coefficient is found by adding the contribution from both walls. Hence, assuming that the body is equidistant from the walls the drag for motion perpendicular to the centerline can be written as

$$\zeta_n = \frac{8\pi\mu\sqrt{r_b}}{\sqrt{h-2r_b}}, \tag{5}$$

where $h$ is the separation between the two walls. Though the normal motion of the rotor is governed by lubrication forces, tangential motions do not induce lubrication. The tangential drag on these systems therefore depends on the full geometry near the walls. Conveniently in the limit $r_b \ll h/2 \ll L$, where $L$ is the arclength, the tangential drag on a cylinder between two walls has been shown to be [43]

$$\zeta_t = \frac{2\pi\mu}{\log(h/r_b)-0.453}. \tag{6}$$

Formally this coefficient does not apply to the gel swimmer as $r_b \sim h/2$. However, as lubrication forces are typically much larger, it is a suitable approximation.

In the resistive force theory formulation $R_\tau$ is found by determining the torque on the fluid generated from a unit angular velocity perpendicular to the planes, $\hat{\mathbf{z}}$. Mathematically this means evaluating

$$R_\tau = \int ds\, \mathbf{r}(s) \times \{[\zeta_t \hat{\mathbf{t}}(s)\hat{\mathbf{t}}(s) + \zeta_n(\mathbf{I} - \hat{\mathbf{t}}(s)\hat{\mathbf{t}}(s))] \cdot (\hat{\mathbf{z}} \times \mathbf{r}(s))\} \tag{7}$$

where $\mathbf{r}(s)$ is the equation for centerline of the slender-body centered around the attachment point and $s$ is the arclength. The power dissipation by fluid motion can therefore be estimated using this formula if a suitable approximation of the shape is available.

Since the spiral rotor periodically changes between a wound loop and a straight rod, we will estimate the viscous power dissipation for both a rotating loop and rotating straight rod. The centerline description for each of these cases are

$$r_{rod}(s) = \{s, 0, 0\}, \tag{8}$$

$$r_{loop}(s) = R\left\{\sin\left(\frac{2\pi s}{L}\right), 1 - \cos\left(\frac{2\pi s}{L}\right), 0\right\}, \tag{9}$$

where $s$ goes from 0 to $L$, $R$ is the radius of the loop and we have written the parametrization in the Cartesian coordinates $\{x, y, z\}$. For these shapes the linearity coefficients become

$$R_\tau^{loop} = \pi R^3 (\zeta_n + 3\zeta_t), \tag{10}$$

$$R_\tau^{rod} = \zeta_n \frac{L^3}{3}. \tag{11}$$

These values correspond to power outputs of

$$P_{loop} \approx 3.75 - 12.3 \times 10^{-16} \text{J s}^{-1},$$

$$P_{rod} \approx 1.6 - 6.9 \times 10^{-14} \text{J s}^{-1},$$

where we have used $R = 25 \mu m$, $\Omega = 0.385$ radians s$^{-1}$, $\mu = 10^{-3} Pa\, s$, $h = 10 \mu m$, $L = 240$ μm and $r_b = 3.2 - 4.85$ μm.

The irradiation energy exerted on the microgel can be estimated from the exposure intensity (1.7 W mm$^{-2}$) and the surface area of the microgel. By assuming the incident angle is 90°, the surface area of the microgel that receives irradiation is roughly 1000 μm². The average input power is therefore $1.7 \times 10^{-3}$ J s$^{-1}$. The efficiency can thus be calculated to be on the order of 10$^{-12}$ - 10$^{-13}$ for loops and 10$^{-11}$ - 10$^{-12}$ for rods. To put this in context, the efficiency of other light powered microscopic devices have been estimated between 10$^{-8}$ - 10$^{-10}$ for thermocapillary devices,[44] 10$^{-11}$ for bacterial devices,[45] and 10$^{-15}$ for direct momentum transfer.[44] The efficiency obtained herein is therefore much more effective than direct momentum transfer techniques and is of a similar magnitude to other light powered systems.

Furthermore, this efficiency can be thought of as a lower bound for the following reasons. (1) The relatively low absorbance (~ 0.12 in the thickness direction) of the microgel object due to small dimension, which means roughly 75% of incident light energy is transmitted through the hydrogel object and lost. (2) Large heat capacity of the gel that requires a significant portion of the energy to be consumed in order to heat the microgel object. (~ $6 \times 10^{-7}$ J in four seconds, see Supporting Information) (3) Constant heat dissipation due to temperature difference between the microgel and surrounding water. Therefore, the efficiency can be improved if the above aspects are taken into consideration, such as further increasing the concentration of AuNRs, and choosing a polymer matrix with a lower volume phase transition temperature so that less energy is required to trigger the deformation.

## 3. Conclusion

We have engineered a micron-scale hydrogel bilayer ribbon that curls upon swelling in water, resulting in a spiral with tunable mechanical responses. The photo-induced heating of the gold nanorods embedded in the microgel ensures a rapid and precise change in temperature within the object and can be utilized to control the swelling state of the PNIPAM network. The resulting expansion/shrinkage of the hydrogel causes the winding and unwinding of the spiral. Energy-transducing nanorods combined with the thermal dissipation capacity of water enabled microgel shape to change out of the thermodynamic equilibrium, as the thermal diffusion processes (heating/cooling) are up to 100 times faster than the mass diffusion processes (volume change).

Light as an energy feed is particularly relevant because the frequency of strobe irradiation now dictates the kinematics of curling deformation and allows non-reciprocal bending and unbending cycles that follow distinct paths, which are required for a net rotational thrust at low *Re*. Using a resistive force formalism, the efficiency of the current system was estimated to be about $10^{-12}$ and is comparable with other light driven systems. Furthermore, we have proposed additional experimental considerations in order to improve the efficiency of our system for future investigations.

The present study opens the new avenue for artificial microswimmers that can be remotely actuated by a near-infrared laser with temporal resolution down to milliseconds and spatial responsiveness in the micrometer range. This design principle may also be extended to program other motile elements, in order to be utilized for biomedical and microfluidic applications, as well as for the development of novel soft micro-robotics.

## 4. Experimental Section

*Synthesis of gold nanorods*: AuNRs were synthesized via a modified seed-mediated method.[46] All chemicals were purchased from Sigma-Aldrich unless otherwise mentioned, and deionized water (0.1 µS cm$^{-1}$, ELGA Purelab-Plus) was used throughout the experiments. To make the seeds, 0.60 mL of freshly prepared ice-cold 0.010 M NaBH$_4$ (99%) aqueous solution was quickly injected to a mixture of water (4.2 ml), 0.20 M Cetyltrimethylammonium bromide (CTAB, 5.0 ml, 99%) and 0.0030 M hydrogen tetrachloroauric(III) acid (HAuCl$_4$, 0.83 mL, 99.9%) under vigorous stirring, and the seed solution was further stirred for 10 min. For the growth solution, 0.20 M CTAB (150 ml), 0.050 M ascorbic acid (3.1 ml, 99%), and 0.0080 M AgNO$_3$ (3.3 ml, 99.99%) were added to 0.0010 M HAuCl$_4$ (150 ml) under stirring. The temperature of growth solution was kept at 25 °C in water bath. 0.875 mL of the seed solution was injected into the growth solution under vigorous stirring for 30 min. Afterwards 0.050 M ascorbic acid (2.0 ml) was added at a flow rate of 0.50 ml h$^{-1}$, followed by 30 min of further stirring. A brownish red solution of AuNRs was then obtained. Centrifugation was carried out to the as-prepared solution at 8000 rpm for 40 min (Centrifuge 5810, Eppendorf) to remove the excess of surfactant. The supernatant was discarded and the precipitated AuNRs were collected and re-suspended in water to a final volume of ca. 20 ml.

*Modification of gold nanorods*: Functional polyethylene glycol (PEG) polymer (HS-PEG-OH, *M*w = 3000 Da, Iris Biotech) was dissolved in ethanol (99.8%) to make a 2.5 mM solution. 10 mL of purified AuNRs was diluted to 100 mL followed by addition of the ethanolic PEG solution (20 mL) under stirring. The solution was then sonicated at 60 °C for 30 min and at 30 − 50 °C for another 3.5 h.[47] Stirring of the solution was then carried out overnight. Subsequently, the solution was extracted by chloroform (120 ml, p.a.) for three times to remove the CTAB and unbound PEG. The aqueous phase after extraction was centrifuged for 3 times, where the supernatant was discarded each time and the precipitate (less than 2 mL) was diluted with DMSO to 45 ml. After the last centrifugation, the residue was collected as a concentrated solution of PEGylated AuNRs in DMSO. UV-Vis sprectrum (V-630, JASCO) shows that the modified AuNRs have an absorption maximum of longitudinal band at 791 nm.[20] Analysis based on transmission electron microscope (Zebra 120, Zeiss) reveals that the AuNRs have an average diameter of 15.4 nm and length of 60.0 nm, i.e. aspect ratio of 3.90.

*Preparation of non-wetting template*: Microgel particles were fabricated using non-wetting molds as template. A cleaned microscope slide was placed on top of a diced silicon master (2 cm × 2 cm, for surface structures see **Figure S1**) produced by photolithography (AMO GmbH), where two pieces of four-layered parafilm (Bemis, total thickness ~ 400 µm) were used as spacers. Perfluoropolyether-urethane dimethacrylate (PFPE, $M_w$ = 2000, Fluorolink MD700, Solvay Solexis) containing 1 wt% Darocure 1173 (Ciba Speciality Chemicals) was then injected into the space until it was fully filled. Subsequently, the PFPE was cured for 20 min under a UV lamp (366 & 254nm, 2 × 4 W, Konrad Benda) in argon atmosphere. After curing, the PFPE film with replicated structures was carefully peeled off from the silicon master and cut into suitable size (6 mm × 6 mm). The flat PFPE film was prepared with the same protocol except that another microscope slide was used instead of silicon master.

*Fabrication of AuNR-laden microgels*: Crosslinker (*N,N′*-Methylenebisacrylamide, BIS, 99%) and photo-initiator (2-Hydroxy-4′-(2-hydroxyethoxy)-2-methylpropiophenone, 98%) were added at a molar amount of 1% relative to *N*-isopropylacrylamide (NIPAM, 97%, recrystallized twice in n-hexane). A typical monomer solution thus contains 57.5 mg of NIPAM, 0.78 mg of BIS, 1.14 mg of photo-initiator, 57.5 µL of AuNR solution, and 40.6 µL of DMSO. The optical density of the solution is 240 at 791 nm (1 cm optical path) due to the presence of AuNRs.

The fabrication procedure of microgel particles is the same as used in our previous reports.[17,20] A home-made press with a quartz window was used for the molding of microgel, and all steps were carried out in a glove box ($O_2$ < 0.2%). The monomer solution (0.5 µL) was first pipetted on the replicated PFPE film, which was then covered by a flat PFPE film. A weight was applied on top of the flat film to generate suitable pressure (~ 260 kPa) to the mold. The pressure ensures that only separate elements were formed. The microgels were cured by 20 min of UV irradiation (366 & 254 nm, 4 W at each wavelength, Konrad Benda). Afterwards the weight was removed, and the mold was peeled off from the flat film. Bi-layered microgels were produced by sputtering a thin gold film (~ 2 nm) on the mold with microgels by a sputter coater (30 mA, 20s, Edwards S150B).

To transfer the microgels, microscope slides (Corning glass) were cut into squares (25 mm × 25 mm) and cleaned with sonication in isopropanol. 0.5 mm thick PDMS film (Sylgard 184, Dow Corning) was cut into square frame with outer dimension of 15 mm × 15 mm and inner dimension of 10 × 10 mm. The cleaned glass slide and the PDMS frame were activated in $O_2$ plasma (200 W, 20 s, 1 mbar, PVP Tepla 100) and then bound together to form an open chamber. Glycerol (15 µL, 99.5%) was pipetted in the chamber and the PFPE mold with microgel particles was placed upside down in the chamber. The whole chamber was then stored in a freezer at - 80 °C overnight. Transfer of the microgels was achieved by peeling the PFPE mold off, where the frozen glycerol served as adhesive. The glycerol was subsequently evaporated at 60°C under vacuum ($1 \times 10^{-2}$ mbar), leaving only microgels in the PDMS chamber. Deionized water was added to re-swell the microgel.

To fabricate the rotor, spiral shaped microgels were mixed with 10 µm non-modified silica microparticles (Creative Diagnostics, US), and the microgels were actuated by modulated NIR laser (0.5 s - 0.5 s) for 5 min for thorough mixing. To fix the microparticles, several layers of round coverslips (diameter: 10 mm) were first glued in the center of a 18 mm × 18 mm coverslip with epoxy adhesive, which was then used to cover the PDMS chamber containing microgels as illustrated in **Figure S2**. In this way, the spiral microgels were confined in the space between the cover slip and the glass substrate. The solid microspheres were thus clamped by the two surfaces, while the attached soft microgels can be actuated. The spiral microgel, which had a microsphere attached to one end, was selected for the current study.

*Actuation and observation of microgels*: The samples were placed on a Peltier stage mounted on an optical microscope (VHZ-100UR, Keyence). A silicon wafer was placed beneath the sample to enhance optical contrast. The temperature of the Peltier stage was controlled with an accuracy of ± 0.1 °C. Near infrared (NIR) laser (808 nm, 2.5 W, Roithner Lasertechnik) was focused to the middle of field of view with an incident angle of roughly 40 °. The elliptical spot has a diameter of around 1 mm, and the laser intensity in the center of the spot is roughly 1.7 W $mm^{-2}$. The laser was modulated with a temporal resolution of 1 ms. For the spiral rotor, videos were recorded by the microscope camera at a frame rate of 27 frames per second. A short pass filter (700 nm, OD 4, Edmund optics) was inserted between the lens and camera to filter scattered laser. The setup is schematically shown in Figure S3.

*Video analysis:* Acquired videos were analyzed with ImageJ (V1.50c). Superimpositions of the images were done using the *Image Calculator* function. Figure 3b is the superimpositions of thresholded images, to which individual colors were given using *Channel Tools*. To acquire the change in shape of the spiral, the images were inverted, and the contour was fitted by *Ridge Detection* (Figure S4a). The coordinates of the fitted curve were then exported and analyzed. The length of the center line can be easily calculated by adding up the lengths between adjacent points, while the tangential angle of the free end was measured by linear fitting of the points at the end. In principle, the tangential angle on any point of the center line can be calculated based on the fitting. As an example, the tangential angle at the half length of the spiral is plotted in Figure S4c, which shows a qualitatively similar shape as the one in Figure 3d.

**Acknowledgements**

We thank Tamas Haraszti for the help with image analysis. H.Z., A.M., and M.M. acknowledge the funding by DFG – SPP 1726 Microswimmer (Project number: 255087333). This project has received funding from the European Research Council (ERC) under the European Union's Horizon 2020 research and innovation program (grant agreement 682754 to E.L.) and ERC - Advanced Grant 695716 to M.M.

**Conflict of Interest**

The authors declare no conflict of interests.

Supporting Information

**A Light-Driven Microgel Rotor**

*Hang Zhang, Lyndon Koens, Eric Lauga, Ahmed Mourran\*, Martin Möller*

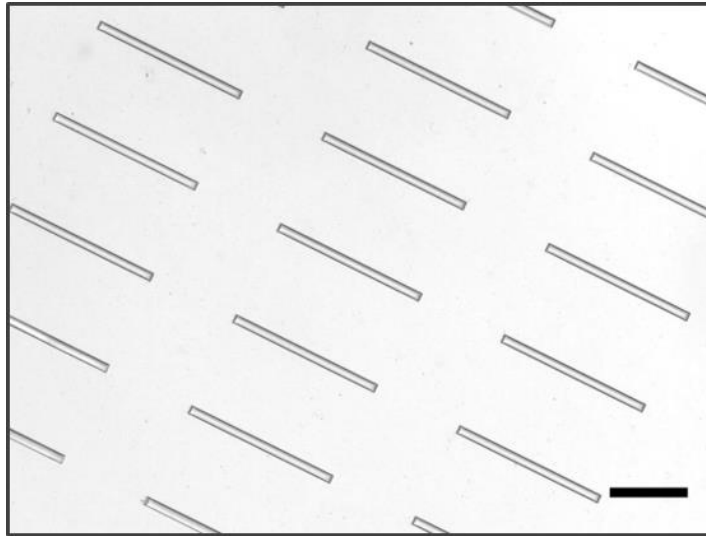

**Figure S1. Optical micrographs of the silicon master**. Surface features: 10 µm × 200 µm rectangles. Etching depth: 5 µm. Scale bar: 100 µm.

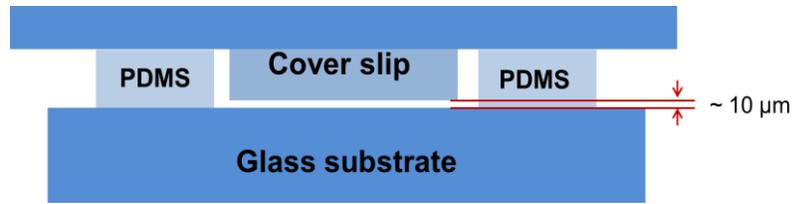

**Figure S2. Illustration of the chamber containing spiral rotor.**

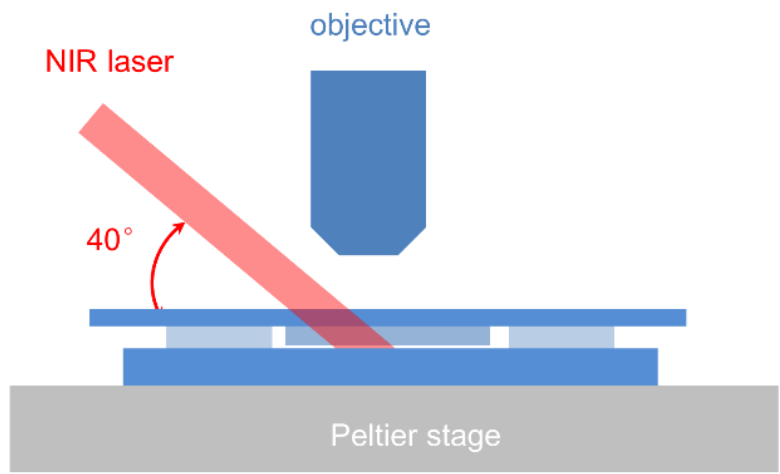

**Figure S3. Schematic drawing of the experimental setup.** The laser was aligned onto the field of view of the microscope objective with an incident angle of 40°.

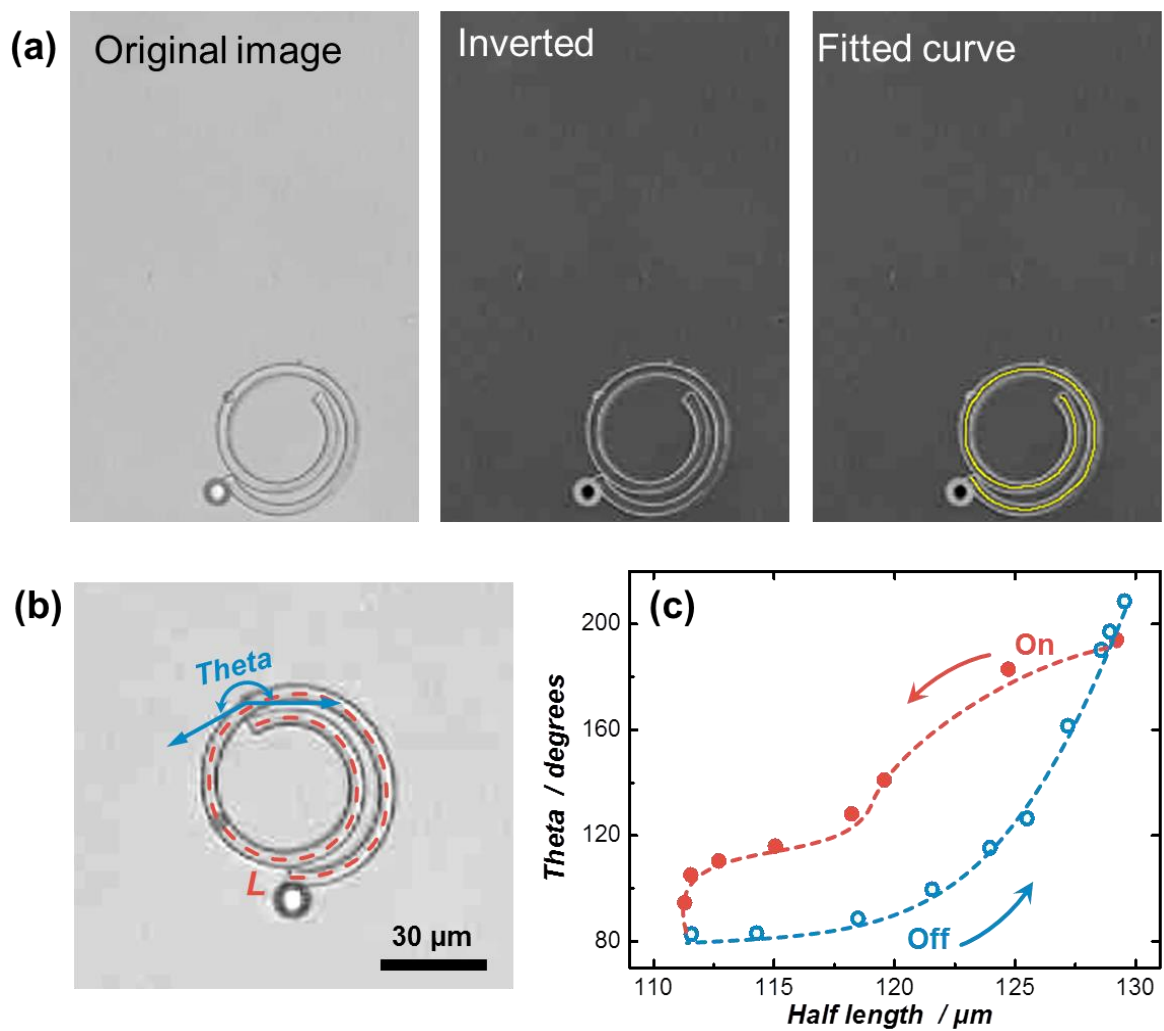

**Figure S4. Image analysis.** (a) Fitting of the spiral microgel using ImageJ. (b) Tangential angle ($\theta$) at the center of the centerline. (c) Change in the $\theta$ with the half length of the center line.

**Calculation of the energy to heat up the microgel from 20 °C to 35 °C**

To calculate the energy required to heat up the microgel, we have adopted the following parameters: temperature increase $\Delta T = 15°C$, heat capacity of the gel $C_{p,gel}= 4.2$ J g$^{-1}$ °C$^{-1}$ (the equivalent of water), weight of the gel $m_{gel} = 10^{-8}$ g. The energy $E$ is calculated following

$$E = \Delta T \times C_{p,gel} \times m_{gel} \sim 6 \times 10^{-7} \text{ J}$$

**Captions for supporting videos**

**Supporting Video 1.** Spiral rotor under 0.1 s – 0.1 s (on - off) modulation of irradiation.

**Supporting Video 2.** Spiral rotor under 0.2 s – 0.2 s (on - off) modulation of irradiation.

**Supporting Video 3.** Spiral rotor under 0.5 s – 0.5 s (on - off) modulation of irradiation.

**Supporting Video 4.** Spiral rotor under 1.0 s – 1.0 s (on - off) modulation of irradiation.